\documentclass[twocolumn,prl,superscriptaddress,showpacs]{revtex4}

\usepackage{euscript,amssymb,amsmath}

\usepackage{graphicx}
\usepackage{epsfig}

\newcommand{\be}[1]{\begin{equation}\label{#1}}
\newcommand{\ee}{\end{equation}}
\newcommand{\ba}[1]{\begin{eqnarray}\label{#1}}
\newcommand{\ea}{\end{eqnarray}}
\newcommand{\rf}[1]{(\ref{#1})}
\newcommand{\nn}{\nonumber}

\begin{document}

\title{Problematic aspect of extra dimensions}

\author{Maxim Eingorn}\email{maxim.eingorn@gmail.com}  \author{Alexander Zhuk}\email{ai_zhuk2@rambler.ru}
\affiliation{Astronomical Observatory and Department of
Theoretical Physics, Odessa National University, Street Dvoryanskaya 2, Odessa 65082, Ukraine}

%
%
%

%
\begin{abstract} We show that in multidimensional Kaluza-Klein models the formula of the perihelion shift is $D\pi m'^2c^2r_g^2/[2(D-2)M^2]$ where $D$ is a
total number of spatial dimensions. This expression demonstrates good agreement with experimental data only in the case of ordinary three-dimensional $(D=3)$ space.
This result does not depend on the size of the extra dimensions. Therefore, considered multidimensional Kaluza-Klein models face a severe problem.
\end{abstract}

\pacs{04.50.-h, 11.25.Mj, 98.80.-k}
\maketitle

\vspace{.5cm}



{\em Introduction}.--- The idea of the multidimensionality of our Universe demanded by the theories of unification of the fundamental interactions is one of the most
breathtaking ideas of theoretical physics. It takes its origin from the pioneering papers by Th. Kaluza and O. Klein \cite{KK} and now
the most self-consistent modern theories of unification such as superstrings, supergravity and M-theory are constructed
in spacetime with extra dimensions (see e.g. \cite{Polchinski}). Different aspects of the idea of the multidimensionality are intensively used in numerous
modern articles.
Therefore, it is very important to suggest experiments which can reveal the extra dimensions. For example, one of the aims of Large Hadronic Collider consists in
detecting of Kaluza-Klein particles which correspond to excitations of the internal spaces (see e.g. \cite{KKparticles}).
On the other hand, if we can show that the existence of the extra dimensions is contrary to observations, then these theories are
prohibited.

It is well known that the perihelion shift of planets is one of important tests of any gravitational theory. There is the significant discrepancy for Mercury between the
measurement value of the perihelion shift and its calculated value using Newton's formalism \cite{Shapiro}.
General relativity is in good agreement with these observations. Obviously, multidimensional gravitational theories should also be in concordance with these experimental
data. To check it, the corresponding estimates were carried out in a number of papers. For example, in \cite{indians}, it was investigated the well known
multidimensional solution \cite{MP} and the authors obtained a negative result. However, this result was clear from the very beginning because the solution \cite{MP}
does not have nonrelativistic Newtonian limit in the case of extra dimensions. Definitely, in solar system such solutions lead to results which are far from the
experimental data. The 5-D soliton metric \cite{soliton} was investigated in \cite{LO} (see also \cite{LOW}). It was found a range of parameters for which the perihelion
shift of Mercury in this model satisfies the observational values. However, our calculations (we will demonstrate it in the extended version of our paper) clearly show
that this range of parameters is quite far from the values which possess the correct nonrelativistic Newtonian limit for a point mass gravitating source. In 5-D
nonfactorizable brane world model, this problem was investigated in \cite{brane}. Here, the model contains one free parameter associated with the bulk Weyl tensor. For
appropriate values of this parameter, the perihelion shift in this model does not contradict to observations. Certainly, this result is of interest and it is necessary
to examine carefully this model to verify the naturalness of the conditions imposed.

In our letter we investigate the perihelion shift of planets in models with an arbitrary number of spatial dimensions $D\ge 3$. We suppose that in the absence of
gravitating masses the metric is a flat one. Gravitating masses (moving or at the rest) perturb this metric and we consider these perturbations in a weak field
approximation. Then we admit that, first, the extra dimensions are compact and have the topology of tori and, second, gravitational potential far away from gravitating
masses tends to nonrelativistic Newtonian limit. All our assumptions are very general and natural. In the case of one gravitating mass at the rest, the obtained metric
coefficients are used to calculate the perihelion shift of a test mass. We demonstrate that this formula depends on a total number of spatial dimensions and its 
application to  Mercury are in good agreement with observations only in ordinary three-dimensional space. It is important to note that this
result does not depend explicitly on the size of the extra dimensions. 
So, we cannot avoid the problem with perihelion shift in a limit of arbitrary small size of the extra dimensions. 
Thus we claim that considered multidimensional Kaluza-Klein models face a severe problem.


{\em Nonrelativistic limit of General Relativity in multidimensional spacetime }.--- To start with, we consider the general form of the multidimensional metric:
\be{1}
ds^2=g_{ik}dx^idx^k=g_{00}\left(dx^0\right)^2+2g_{0\alpha}dx^0dx^{\alpha}+g_{\alpha\beta}dx^{\alpha}dx^{\beta}\, ,
\ee
where the Latin indices $i,k = 0,1,\ldots ,D$ and the Greek indices  $\alpha ,\beta = 1,\ldots ,D$. $D$ is the total number of
spatial dimensions.
We make the natural assumption that in the case of the absence of matter sources the spacetime is Minkowski spacetime: $g_{00}=\eta_{00}=1$,
$g_{0\alpha}=\eta_{0\alpha}=0$, $g_{\alpha\beta}=\eta_{\alpha\beta}=-\delta_{\alpha\beta}$. At the same time, the extra dimensions
may have the topology of tori.
In the presence of matter, the metric is not Minkowskian one and we will investigate it in the weak field limit. It means that the
gravitational field is weak and velocities of test bodies are small compared with the speed of light $c$. In this case the metric is only
slightly perturbed from its flat spacetime value:
\be{2}
g_{ik}\approx\eta_{ik}+h_{ik}\, ,
\ee
where $h_{ik}$ are the corrections of the order $1/c^2$. In particular, $h_{00} \equiv 2\varphi /c^2$. Later we will demonstrate that $\varphi $ is nonrelativistic
gravitational potential. It can be shown e.g. by comparing nonrelativistic and relativistic actions for the point mass particle \cite{Landau}. To get other correction
terms up to the same order $1/c^2$, we should consider multidimensional Einstein equation
\be{3}
R_{ik}=\frac{2S_D\tilde G_{\mathcal{D}}}{c^4}\left(T_{ik}-\frac{1}{D-1}g_{ik}T\right)\, ,
\ee
where $S_D=2\pi^{D/2}/\Gamma (D/2)$ is the total solid angle (surface area of the $(D-1)$-dimensional sphere of unit radius), $\tilde G_{\mathcal{D}}$ is the
gravitational constant in the $(\mathcal{D}=D+1)$-dimensional spacetime and the energy-momentum tensor of $N$ point mass particles is
\be{4} T^{ik}=\sum\limits_{p=1}^Nm_p\left[(-1)^Dg\right]^{-1/2}\frac{dx^i}{dt}\frac{dx^k}{dt}\frac{cdt}{ds}\delta({\bf r}-{\bf r}_p)\, , \ee
where $m_p$ is a rest mass and ${\bf r}_p$ is a radius vector of the $p$-th particle respectively. The rest mass density is
\be{5}
\rho=\sum\limits_{p=1}^Nm_p\delta({\bf r}-{\bf r}_p)\, .
\ee
Holding in the left hand and right hand sides of Eq. \rf{3} terms of the order $1/c^2$ we obtain the following equations:
\ba{6}
\triangle h_{00}&=&\frac{2S_DG_{\mathcal{D}}}{c^2}\rho\, ,\quad \triangle h_{0\alpha}=0\, , \nn \\
\triangle h_{\alpha\beta}&=&\frac{1}{D-2}\cdot\frac{2S_DG_{\mathcal{D}}}{c^2}\rho\delta_{\alpha\beta}\, ,
\ea
where $\triangle = \delta^{\alpha\beta}\partial^2/\partial x^{\alpha}\partial x^{\beta}$ is $D$-dimensional Laplace operator and
$G_{\mathcal{D}}=[2(D-2)/(D-1)]\, \tilde G_{\mathcal{D}}$. Substitution of $h_{00} = 2\varphi /c^2$ into above equation for $h_{00}$
demonstrates that $\varphi $ satisfies $D$-dimensional Poisson equation:
\be{7}
\triangle \varphi = S_DG_{\mathcal{D}}\rho\, .
\ee
Therefore, $\varphi$ is nonrelativistic gravitational potential. From Eqs. \rf{6} we obtain
\be{8} h_{0\alpha}=0\, ,\quad h_{\alpha\beta}=\frac{1}{D-2}\cdot h_{00}\delta_{\alpha\beta}=\frac{1}{D-2}\cdot\frac{2\varphi}{c^2}\delta_{\alpha\beta}\, . \ee
It is worth of noting that the relation $h_{\alpha\beta}/h_{00}=[1/(D-2)]\delta_{\alpha\beta}$ can be also obtained from the corresponding equations in
\cite{MP,CDandMP}.

Now, we want to keep in metric \rf{1} the terms up to the order $1/c^2$. Because the coordinate $x^0=ct$ contains $c$, it means that in $g_{00}$ and $g_{0\alpha}$ we
should keep correction terms up to the order $1/c^4$ and $1/c^3$ respectively and to leave $g_{\alpha\beta}$ without changes in the form $g_{\alpha\beta} \approx
\eta_{\alpha\beta}+h_{\alpha\beta}$ with $h_{\alpha\beta}$ from Eq. \rf{8}. Holding in the left hand and right hand sides of $00$ and $0\alpha$ components of Einstein
equation \rf{3} terms up to the order $1/c^4$ and $1/c^3$ respectively, we obtain after a long but obvious calculations the required correction terms:
\ba{9}
g_{00}&\approx& 1+\frac{2}{c^2}\varphi({\bf r})+\frac{2}{c^4}\varphi^2({\bf r})+\frac{2}{c^4}\sum\limits_{p=1}^N\varphi'_p\varphi'({\bf r}-{\bf
r}_p)\nn \\
&+&\frac{D}{D-2}\cdot\frac{1}{c^4}\sum\limits_{p=1}^Nv_p^2\varphi'({\bf r}-{\bf r}_p)\, , \ea
\be{10}
g_{0\alpha}\approx-\frac{2(D-1)}{D-2}\cdot\frac{1}{c^3}\sum\limits_{p=1}^Nv_{p\alpha}\varphi'({\bf r}-{\bf r}_p)-\frac{1}{c^3}\frac{\partial^2f}{\partial
t\partial x^{\alpha}}\, .
\ee
Here, $v_p^{\alpha}=dx_p^{\alpha}/dt$ is $\alpha$-component of the velocity of the $p$-th particle, $\varphi'_p$ is potential of the gravitational field in a point with
radius vector ${\bf r}_p$ produced by all particles, except for $ p $-th, $\varphi'({\bf r}-{\bf r}_p)$ is potential of the gravitational field of $ p $-th particle
satisfying the Poisson equation $\triangle\varphi'=S_DG_{\mathcal{D}} m_p\delta({\bf r}-{\bf r}_p)$ (therefore, $\varphi({\bf r})=\sum_{p=1}^N\varphi'({\bf r}-{\bf
r}_p)$) and function $f({\bf r})$ satisfies equation $\triangle f=\varphi({\bf r})$. It is worth of noting that the radius vector ${\bf r}_p$ of the $ p $-th particle
may depend on time $t$. Eqs. \rf{9} and \rf{10} generalize the known formulas (see \S $\,$ 106 in \cite{Landau}) to an arbitrary number of dimensions $D\ge 3$. In the
case of one gravitating particle at the rest at the origin of coordinates, the metric coefficients have the form
\ba{11}
g_{00}&\approx& 1+\frac{2}{c^2}\varphi({\bf r})+\frac{2}{c^4}\varphi^2({\bf r}),\ \ \ g_{0\alpha}\approx0,\nn \\
g_{\alpha\beta}&\approx& -\left(1-\frac{1}{D-2}\cdot\frac{2}{c^2}\varphi({\bf r})\right)\delta_{\alpha\beta}\, .
\ea

As we have noted above, we assume that the internal space is compact and has the topology of tori. For this topology, and with the boundary condition that at infinitely
large distances from the gravitating body potential must go to the Newtonian expression, we can find the exact solution of the Poisson equation \rf{7} \cite{EZ1,EZ2}.
The boundary condition requires that the multidimensional and Newtonian gravitational constants are connected by the following condition: $S_D G_{\mathcal{D}}/V =4\pi
G_N$ where $V$ is the volume of the internal space. Assuming that we consider gravitational field of the gravitating mass $m$ at distances much greater than periods of
tori, we can restrict ourselves by the zero Kaluza-Klein mode. For example, this approximation is very well satisfied for the planets of the solar system because the
inverse-square law experiments show that the extra dimensions in Kaluza-Klein models should not exceed submillimeter scales \cite{new} (see however \cite{EZ1,EZ2} for
models with smeared extra dimensions where Newton's law preserves its shape for arbitrary distances). Then, the gravitational potential reads
\be{12}
\varphi({\bf r})\approx -\frac{G_N m}{r_3} = -\frac{r_gc^2}{2r_3}\, ,
\ee
where $r_3$ is the length of a radius vector in three-dimensional space and we introduced three-dimensional Schwarzschild radius $r_g=2G_Nm/c^2$.

{\em Perihelion shift}.--- In the approximation \rf{12}, the covariant components of the metric \rf{11} take the form
\ba{13}
g_{00}&\approx& 1-\frac{r_g}{r_3}+\frac{r_g^2}{2r_3^2},\ \ \ g_{0\alpha}\approx0,\nn \\
g_{\alpha\beta}&\approx&-\left(1+\frac{1}{D-2}\cdot \frac{r_g}{r_3}\right)\delta_{\alpha\beta}\, .
\ea

Let us consider now the motion of a test body of mass $m'$ in a gravitational field described by Eqs. \rf{13}. The Hamilton-Jacobi equation
\be{14}
g^{ik}\frac{\partial S}{\partial x^i}\frac{\partial S}{\partial x^k}-m'^2c^2=0
\ee
reads
\ba{15}
&{}&\frac{1}{c^2}\left(1+\frac{r_g}{r_3}+\frac{r_g^2}{2r_3^2}\right)\left(\frac{\partial S}{\partial t}\right)^2\nn \\
&-&\left(1-\frac{1}{D-2}\cdot\frac{r_g}{r_3}\right)\left(\frac{\partial S}{\partial
r_3}\right)^2\nn\\
&-&\frac{1}{r_3^2}\left(1-\frac{1}{D-2}\cdot\frac{r_g}{r_3}\right)\left(\frac{\partial S}{\partial \psi}\right)^2\nn \\
&-&\left(1-\frac{1}{D-2}\cdot\frac{r_g}{r_3}\right)\left[\left(\frac{\partial S}{\partial x^4}\right)^2+...+\left(\frac{\partial S}{\partial x^D}\right)^2\right]\nn \\
&-& m'^2c^2\approx 0\, ,
\ea
where we use spherical coordinates $(r_3,\theta,\psi)$ in three-dimensional space and consider the motion of a test body in the orbital plane $\theta=\pi/2$. We
investigate this equation by separation of variables, considering the action in the form $S=-E't+M\psi+S_{r_3}(r_3)+S_4\left(x^4\right)+...+S_D\left(x^D\right)$. Here,
$E'\approx m'c^2+E$ is the energy of the test body, which includes the rest energy $m'c^2$ and nonrelativistic energy $E$. Substituting this expression for the action
$S$ in the formula \rf{15}, we obtain an expression for $(dS_{r_3}/dr_3)^2$  holding there the members up to the order $1/c^2$. Integrating the square root of this
expression with respect to $r_3$, we finally get $S_{r_3}$ in the following form:
\ba{16}
S_{r_3}&\approx&\int \left[\left(2m'E-\left(p_4^2+...+p_D^2\right)+\frac{E^2}{c^2}\right)\right.\nn \\
&+&\frac{1}{r_3}\left(m'^2c^2r_g+\frac{2(D-1)}{D-2}\, m'Er_g\right)\nn\\
&-&\left.\frac{1}{r_3^2}\left(M^2-\frac{Dm'^2c^2r_g^2}{2(D-2)}\right)
\right]^{1/2}dr_3\, ,
\ea
where $p_{\alpha} = \partial S/\partial x^{\alpha}=dS_{\alpha}/dx^{\alpha}\; (\alpha = 4,\ldots ,D)$ are the components of momentum of the test body in the extra dimensions.
If the gravitating and test masses are localized on the same brane then these components are equal to zero.

The trajectory of the test body is defined by the equation $\partial S/\partial M=\psi+\partial S_{r_3}/\partial M=\mathrm{const}$. Let now the Sun be the gravitating
mass, and the planets of the solar system be the test bodies. Then, the change of the angle during one revolution of a planet on an orbit is
\be{17}
\Delta\psi=-\frac{\partial}{\partial M}\Delta S_{r_3}\, ,
\ee
where $\Delta S_{r_3}$ is the corresponding change of $S_{r_3}$. It is well known that small relativistic correction $\delta \equiv \frac{Dm'^2c^2r_g^2}{2(D-2)}$ to $M^2$
in Eq. \rf{16} results in the perihelion shift. Expanding $S_{r_3}$ in powers of this correction, we obtain
\be{18}
S_{r_3}\approx S_{r_3}^{(0)}-\frac{Dm'^2c^2r_g^2}{4(D-2)M}\frac{\partial S_{r_3}^{(0)}}{\partial M}\, ,
\ee
where $S_{r_3}^{(0)}=S_{r_3}(\delta =0)$. From this equation we obtain
\be{19}
\Delta\psi\approx2\pi+\frac{D\pi m'^2c^2r_g^2}{2(D-2)M^2}\, ,
\ee
where we took into account $-\partial \Delta S_{r_3}^{(0)}/\partial M=\Delta\psi^{(0)}=2\pi$. Therefore, the second term in \rf{19} gives the required formula for the
perihelion shift in our multidimensional case.

It make sense to apply this formula to Mercury because in the solar system it has the most significant discrepancy between the measurement value of the perihelion shift
and its calculated value using Newton's formalism. The observed discrepancy is $43.11\pm 0.21$ arcsec per century \cite{Shapiro}. This missing value is usually explained
by the relativistic effects of the form of \rf{19}.
However, only in three-dimensional case $D=3$ Eq. \rf{19} gives the satisfactory result $42.94''$ which is within the measurement accuracy. For $D=4$ and $D=9$ models we
obtain $28.63''$ and $18.40''$, respectively, which is very far from the observable value. It is worth of noting that this result does not depend on the size of the
extra dimensions (up to the applicability of the approximation \rf{12}). It is not difficult to generalize our consideration to the case of the gravitating mass moving
in the extra dimensions (but at the rest with respect to our three dimensions). This generalization does not change Eq. \rf{19}.


{\em Summary}.--- We have investigated the perihelion shift of planets for multidimensional models with compact internal spaces in the form of tori. We have found that
the obtained formula for the perihelion shift depends on the total number of spatial dimensions. Our estimates show that only three-dimensional case $D=3$ is in good
agreement with the experimental data and all multidimensional cases $D>3$ contradict observations. This result does not depend on the size of the extra dimensions.
Therefore, considered multidimensional Kaluza-Klein models face a severe problem.


\indent \indent We want to thank Uwe G\"uenther for useful discussion. This work was supported in part by the "Cosmomicrophysics" programme of the Physics and Astronomy
Division of the National Academy of Sciences of Ukraine.



\end{document}